\documentclass[useAMS,usenatbib]{mn2e}
\usepackage{txfonts}
\usepackage{natbib}
\usepackage[all]{xy}
\usepackage[dvips]{graphicx}

\def\del#1{{}}

\newcommand{\ltsima}{$\; \buildrel < \over \sim \;$}
\newcommand{\lsim}{\lower.5ex\hbox{\ltsima}}
\newcommand{\gtsima}{$\; \buildrel > \over \sim \;$}
\newcommand{\gsim}{\lower.5ex\hbox{\gtsima}}
\newcommand{\bra}{\langle}
\newcommand{\ket}{\rangle}

\newcommand{\dd}{\mathrm{d}}

\newcommand{\veck}{\bmath{k}}
\newcommand{\vecp}{\bmath{p}}
\newcommand{\vecq}{\bmath{q}}

\title[linear and nonlinear iSW-effect]
{Parameter estimation biases due to contributions from the Rees-Sciama effect to the integrated Sachs-Wolfe spectrum}
\author[B.M. Sch{\"a}fer, A.F. Kalovidouris, L. Heisenberg]
{Bj{\"o}rn Malte Sch\"afer\thanks{e-mail: spirou@ita.uni-heidelberg.de}$^{1,2}$, Angelos Fotios Kalovidouris$^2$ and Lavinia Heisenberg$^{3,4}$\\
$^1$Institut d'Astrophysique Spatiale, Universit{\'e} Paris XI, b{\^a}timent 120-121, Centre universitaire d'Orsay, 91440 Orsay CEDEX, France\\
$^2$Astronomisches Recheninstitut, Zentrum f{\"u}r Astronomie, Universit{\"a}t Heidelberg, M{\"o}nchhofstra{\ss}e 12, 69120 Heidelberg, Germany\\
$^3$Institut f{\"u}r theoretische Astrophysik, Zentrum f{\"u}r Astronomie, Universit{\"a}t Heidelberg, Albert-Ueberle-Stra{\ss}e 2, 69120 Heidelberg, Germany\\
$^4$D{\'e}partement de physique th{\'e}orique, Universit{\'e} de Gen{\`e}ve, 24, quai Ernest Ansermet, 1211 Gen{\`e}ve, Switzerland}

\begin{document}
\pagerange{\pageref{firstpage}--\pageref{lastpage}}
\pubyear{2008}
\maketitle
\label{firstpage}

\begin{abstract}
The subject of this paper is an investigation of the nonlinear contributions to the spectrum of the integrated Sachs-Wolfe (iSW) effect. We derive the corrections to the iSW-auto spectrum and the iSW-tracer cross-spectrum consistently to third order in perturbation theory and analyse the cumulative signal-to-noise ratio for a cross-correlation between the PLANCK and EUCLID data sets as a function of multipole order. We quantify the parameter sensitivity and the statistical error bounds on the cosmological parameters $\Omega_m$, $\sigma_8$, $h$, $n_s$ and $w$ from the linear iSW-effect and the systematical parameter estimation bias due to the nonlinear corrections in a Fisher-formalism, analysing the error budget in its dependence on multipole order. Our results include: $(i)$ the spectrum of the nonlinear iSW-effect can be measured with $0.8\sigma$ statistical significance, $(ii)$ nonlinear corrections dominate the spectrum starting from $\ell\simeq10^2$, $(iii)$ an anticorrelation of the CMB temperature with tracer density on high multipoles in the nonlinear regime, $(iv)$ a much weaker dependence of the nonlinear effect on the dark energy model compared to the linear iSW-effect, $(v)$ parameter estimation biases amount to less than $0.1\sigma$ and weaker than other systematics.
\end{abstract}


\section{Introduction}
The integrated Sachs-Wolfe (iSW) effect \citep{1967ApJ...147...73S, 1994PhRvD..50..627H, 2002PhRvD..65h3518C}, which refers to the frequency change of cosmic microwave background (CMB) photons if they cross time evolving gravitational potentials, is a direct probe of dark energy because it vanishes in cosmologies with $\Omega_m=1$ \citep{1996PhRvL..76..575C}. By now, it has been detected with high significance with a number of different tracer objects \citep{2003ApJ...597L..89F, 2004Natur.427...45B, 2004ApJ...608...10N, 2005PhRvD..72d3525P, 2006PhRvD..74f3520G, 2006PhRvD..74d3524P, 2006MNRAS.365..171G, 2006MNRAS.372L..23C, 2006MNRAS.365..891V, 2007MNRAS.377.1085R, 2007MNRAS.376.1211M, 2008arXiv0801.4380G}, and derived parameter constraints provide support for a $\Lambda$CDM cosmology.

Contrarily, the nonlinear iSW-effect, or Rees-Sciama (RS) effect \citep{rees_sciama_orig, 1996ApJ...460..549S, 2006MNRAS.369..425S} is difficult to detect and shows only a weak signal amouting to $<2\sigma$ in the spectrum \citep{2002PhRvD..65h3518C} or up to $0.8\sigma$ in the bispectrum \citep{2008MNRAS.388.1394S}. The cross-correlation with weak lensing has been shown to be feasible, but weak with current surveys \citep{2008ApJ...676L..93N}. In comparison to the linear iSW-effect, the RS-effect shows a flatter spectral dependence and dominates the signal at higher multipoles exceeding $\ell\gsim100$. Analytical, perturbative derivations agree well with the results from $n$-body simulations \citep{1996ApJ...463...15T, 2008arXiv0809.4488C, 2009arXiv0905.2408S, 2010arXiv1003.0974C}. The non-Gaussianities introduced into the CMB by the nonlinear RS-effect are very weak \citep[although the first two papers work in the context of a SCDM-cosmology, their results are still applicable to $\Lambda$CDM]{1995ApJ...453....1M, 1995ApJ...454..552M, 1999PhRvD..59j3001S,1999PhRvD..59j3002G}. The RS-effect from the local Universe has been found to amount to $\sim2\umu$K in the most massive structures \citep{2007A&A...476...83M} forming in a constraint realisation.

The topic of this paper is the contamination of the iSW-spectrum by the nonlinear RS-spectrum at intermediate multipoles: In a measurement of the linear iSW-effect, nonlinear contributions will alter the shape of the observed spectrum and can affect the estimation of cosmological parameters by introducing estimation biases. We investigate dependence of parameter accuracy as well as the parameter estimation bias as a function of maximum multipole order considered. Specifically, we use a Fisher-matrix approach to quantify the statistical and systematical errors, analyse the error budget as a function of multipole order and derive the optimal maximum multipole moment which minimises the combined error for individual parameters. The nonlinear iSW-effect is the most important contaminant at intermediate multipoles, with the kinetic Sunyaev-Zel'dovich effect starting to dominate at higher multipoles above thousand.

After summarising key formul{\ae} describing structure formation in dark energy cosmologies in Sect.~\ref{sect_homogeneous}, we introduce line of sight expressions of the two relevant observables in Sect.~\ref{sect_channels}. We carry out a perturbative expansion of the source fields to third order in Sect.~\ref{sect_perturbation} and derive the spectrum $C_{\tau\gamma}(\ell)$ between iSW-temperature perturbation $\tau$ and the galaxy density $\gamma$ to third order in Sect.~\ref{sect_isw}. We quantify the degeneracies between the cosmological parameters using a Fisher-matrix analysis in Sect.~\ref{isw_fisher} and extend this formalism to describe the parameter estimation bias in Sect.~\ref{isw_bias}. A summary of our results is compiled in Sect.~\ref{sect_summary}. 

As cosmologies, we consider spatially flat homogeneous dark energy models with constant dark energy equation of state, and with Gaussian adiabatic initial conditions in the cold dark matter field. Specific parameter choices for the $w$CDM-fiducial model in the Fisher-matrix analysis are $H_0=100h \:\mathrm{km}/s/\mathrm{Mpc}$ with $h=0.72$, $\Omega_m=0.25$, $\Omega_b=0.04$, $\sigma_8=0.8$, $w=-0.9$ and $n_s=1$, with constant unit bias for the tracer galaxy population.

\section{Cosmology and structure formation}\label{sect_homogeneous}

\subsection{Dark energy cosmologies}
In a spatially flat dark energy cosmology with a constant dark energy equation of state parameter $w$, the Hubble function $H(a)=\dd\ln a/\dd t$ is given by
\begin{equation}
\frac{H^2(a)}{H_0^2} = \frac{\Omega_m}{a^{3}} + \frac{1-\Omega_m}{a^{3(1+w)}}.
\end{equation}
The value $w\equiv -1$ corresponds to the cosmological constant $\Lambda$. The conformal time, which is related to the cosmic time $t$ by the differential $\dd\eta=\dd t/a$, follows directly from the definition of the Hubble function,
\begin{equation}
\eta = \int_a^1\dd a\: \frac{1}{a^2H(a)},
\end{equation}
in units of the Hubble time $t_H=1/H_0$. Correspondingly, the definition of the comoving distance $\chi$ is given by $\chi=c\eta$ with the speed of light $c$.

\subsection{CDM power spectrum}
A common parameterisation for the CDM power spectrum is $P(k)\propto k^{n_s} T^2(k)$ for describing the Gaussian fluctuation statistics of the homogeneous and isotropic cosmic density field $\delta$, 
\begin{equation}
\bra\delta(\veck)\delta(\veck^\prime)^*\ket = (2\pi)^3\delta_D(\veck-\veck^\prime)P(k)
\end{equation}
According to \citet{1986ApJ...304...15B}, a convenient fit to the CDM transfer function $T(k)$ is
\begin{displaymath}
T(q) = \frac{\ln(1+2.34q)}{2.34q}\left(1+3.89q+(16.1q)^2+(5.46q)^3+(6.71q)^4\right)^{-\frac{1}{4}},
\end{displaymath}
where the wave vector $q$ is given in units of the shape parameter $\Gamma\simeq\Omega_m h$. $P(k)$ is normalised to the value $\sigma_8$ on the scale $R=8~\mathrm{Mpc}/h$,
\begin{equation}
\sigma_R^2 = \frac{1}{2\pi^2}\int\dd k\: k^2 W^2(kR) P(k),
\end{equation}
with a Fourier-transformed spherical top-hat $W(x)=3j_1(x)/x$ as the filter function. $j_\ell(x)$ denotes the spherical Bessel function of the first kind of order $\ell$ \citep{1972hmf..book.....A}. \citet{2009arXiv0905.2408S} found that nonlinear effects in the biasing model amount to $\sim10\%$, but for simplicity, we assume a linear, local, non-evolving and scale-independent biasing scheme,
\begin{equation}
\frac{\Delta n}{n} = \frac{\Delta\rho}{\rho},
\end{equation}
and relate fluctuations $\Delta n$ in the spatial number density $n$ of galaxies directly to the dark matter overdensity $\delta=\Delta\rho/\rho$.

\subsection{Structure growth in dark energy cosmologies}
The linearised structure formation equations, i.e. the continuity, Jeans and Poisson equations, can be combined to the growth equation \citep{1998ApJ...508..483W, 1997PhRvD..56.4439T, 2003MNRAS.346..573L},
\begin{equation}
\frac{\dd^2}{\dd a^2}D_+
+\frac{1}{a}\left(3+\frac{\dd\ln H}{\dd\ln a}\right)\frac{\dd}{\dd a}D_+ = 
\frac{3}{2a^2}\Omega_m(a)D_+(a).
\label{eqn_growth}
\end{equation}
whose solution $D+(a)$ describes the homogeneous growth of the density field, $\delta(\bmath{x},a)=D_+(a)\delta(\bmath{x},1)$. In the standard cold dark matter (SCDM) cosmology with $\Omega_m=1$ and $3+\dd\ln H/\dd\ln a=\frac{3}{2}$, this solution is easily derived to be $D_+(a)=a$. This motivates the choice $D_+(0)=0$ and $\dd/\dd a D_+(0)=1$ for the initial conditions, due to matter domination at early times. The second solution $D_-(a)=1/a$ decays rapidly and has no influence on the late-time iSW-effect.

\section{Observables: iSW-effect and tracers}\label{sect_channels}

\subsection{iSW-temperature perturbation}
The iSW-effect is caused by gravitational interactions of CMB photons with time-evolving potentials $\Phi$. The fractional perturbation $\tau$ of the CMB temperature $T_\mathrm{CMB}$ is given by \citep{1967ApJ...147...73S, rees_sciama_orig}
\begin{equation}
\tau 
= \frac{\Delta T}{T_\mathrm{CMB}} 
= -\frac{2}{c^3}\int_0^{\chi_H}\dd\chi\: a^2 H(a) \frac{\partial\Phi}{\partial a},
\label{eqn_sachs_wolfe}
\end{equation}
The gravitational potential $\Phi$ is a solution to the comoving Poisson equation,
\begin{equation}
\Delta\Phi = \frac{3H_0^2\Omega_m}{2a}\delta.
\end{equation}
Substituting into the line of sight expression for the linear iSW-effect $\tau$ (integrating along a straight line and using the flat-sky approximation) yields
\begin{equation}
\tau = 
\frac{3\Omega_m}{c}\int_0^{\chi_H}\dd\chi\: 
a^2 H(a)\:\frac{\dd}{\dd a}\left(\frac{D_+}{a}\right)\:\frac{\Delta^{-1}}{\chi_H^2}\delta,
\end{equation}
where the inverse Laplace operator $\Delta^{-1}/\chi_H^2$ solves for the potential:
\begin{equation}
\varphi\equiv\frac{\Delta^{-1}}{\chi_H^2}\delta.
\end{equation}
The square of the Hubble distance $\chi_H=c/H_0$ makes the differential operator dimensionless.

\subsection{Galaxy density as a large-scale structure tracer}
The projected galaxy density $\gamma$ can be related to the CDM density $\delta$ via
\begin{equation}
\gamma = \int_0^{\chi_H}\dd\chi\:p(z)\frac{\dd z}{\dd\chi} D_+(\chi)\:\delta,
\end{equation}
where $p(z)\dd z$ is the redshift distribution of the surveyed galaxy sample, rewritten in terms of the comoving distance $\chi$. We use the redshift distribution of the main galaxy sample of EUCLID \citep{2008arXiv0802.2522R}, which will $f_\mathrm{sky}=0.5$ of the sky with a median redshift of $z_\mathrm{med}=0.9$ \citep{2008arXiv0802.0983D}. We use the parameterisation proposed by \citet{1995MNRAS.277....1S}
\begin{equation}
p(z)\dd z = p_0\left(\frac{z}{z_0}\right)^2\exp\left(-\left(\frac{z}{z_0}\right)^\beta\right)\dd z
\quad\mathrm{with}\quad \frac{1}{p_0}=\frac{z_0}{\beta}\Gamma\left(\frac{3}{\beta}\right),
\end{equation}
for $p(z)\dd z$, with $z_0=0.64$. We assume a constant bias of $b=1$, which we absorb into the normalisation $\sigma_8$ of the power spectra.

\section{Perturbative corrections}\label{sect_perturbation}
\begin{figure}
\resizebox{\hsize}{!}{\includegraphics{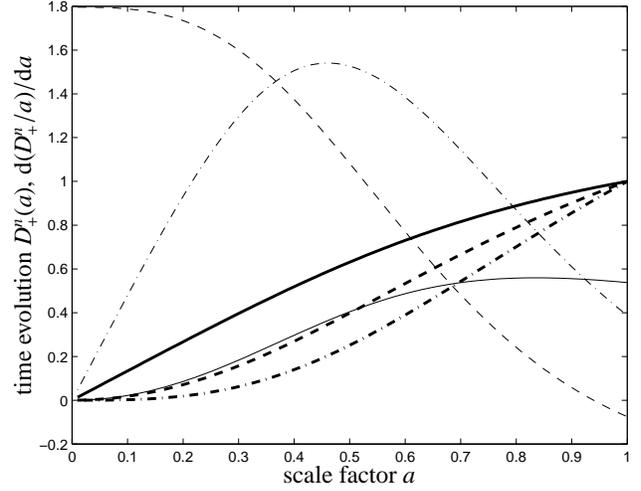}}
\caption{Time evolution of the source term  $D_+^n(a)$ for the density field (thick line) and the modulus of $\dd (D_+^n/a)/\dd a$ for the iSW-effect (thin line), for the linear order $n=1$ (solid line) and the nonlinear corrections $n=2$ (dashed line) and $n=3$ (dash-dotted line), with $\Lambda$CDM as the cosmological model.}
\label{figure_time_evolution}
\end{figure}

For a consistent derivation of the iSW-spectrum including corrections due to the nonlinearly evolved source fields one needs to carry out a perturbative expansion to third order,
\begin{equation}
\delta(\bmath{x},a) \simeq \sum_{n=1}^3 D_+^n(a)\delta^{(n)}(\bmath{x}) + \mathcal{O}(\delta^4).
\end{equation}
The linearity of the Poisson equation conserves the perturbative series,
\begin{equation}
\dot{\varphi}(\bmath{x},a) \simeq \sum_{n=1}^3 \frac{\dd}{\dd a}\frac{D_+^n}{a}\varphi^{(n)} + \mathcal{O}(\varphi^4),
\end{equation}
and suggests that the time derivative of the potential is $\propto \dd (D_+^n/a)/\dd a$. In perturbation theory, the second and third order corrections to the density field are given by 
\begin{equation}
\delta^{(2)}(\bmath{k}) =
\int\frac{\dd^3p}{(2\pi)^3}\:
M_2(\veck-\vecp,\vecp)\delta(\vecp)\delta(\left|\veck-\vecp\right|),
\end{equation}
\begin{equation}
\delta^{(3)}(\bmath{k}) = 
\int\frac{\dd^3p}{(2\pi)^3}\int\frac{\dd^3q}{(2\pi)^3}\:
M_3(\vecp,\vecq,\veck-\vecp-\vecq)\delta(\vecp)\delta(\vecq)\delta(\left|\veck-\vecp-\vecq\right|),
\end{equation}
where the mode coupling functions $M_2(\bmath{p},\bmath{q})$ and $M_3(\bmath{p},\bmath{q},\bmath{r})$ \citep[see][]{1995PhR...262....1S, 2002PhR...367....1B} are a consequence of the inhomogeneous growth and introduce non-Gaussianities in the evolved density field. The power spectrum $P_{\delta\delta}^{(11)}(k) = P(k)$ of the density field thus acquires the corrections
\begin{eqnarray}
P_{\delta\delta}^{(22)}(k) & = & 2\int\frac{\dd^3p}{(2\pi)^3}\:M_2(\veck-\vecp,\vecp)^2P(\left|\veck-\vecp\right|)P(\left|\vecp\right|),\\
P_{\delta\delta}^{(13)}(k) & = & 3\int\frac{\dd^3p}{(2\pi)^3}\:M_3(\veck,\vecp,-\vecp)P(k)P(\left|\vecp\right|).
\end{eqnarray}
In the computation of these corrections, the cylindrical symmetry of the kernels $M_2$ and $M_3$ can be taken advantage of, reducing to a twofold integration $2\pi p^2\dd p\dd\mu$ with $\mu$ being the cosine of the angle between $\veck$ and $\vecp$. Fig.~\ref{figure_time_evolution} shows the time evolution of the source fields, i.e. the growth function $D_+^n(a)$ for the density field, and the time derivative of $D_+^n(a)/a$ for the iSW-source field, both up to perturbative order $n=3$. While the growth functions show a similar behaviour in higher order, the derivatives are qualitatively very different. The evalutations of the integrals is done in a coordinate system whose $p_z$-axis is parallel to $k_z$. The nonlinear corrections to the CDM spectrum $P(k)$ are shown in Fig.~\ref{figure_nonlinear}. 

An interesting peculiarity of the nonlinear RS-effect in comparison to the linear iSW-effect is worth mentioning: Whereas in SCDM-cosmologies the iSW-effect vanishes due to $D_+(a)\equiv a$ and is nonzero in dark energy cosmologies, the RS-effect is strongest in SCDM and weaker in dark energy cosmologies, at least at the low redshifts we observe, where $D_+(a)\simeq a^\alpha$ with $\alpha<1$. Furthermore, the cross-spectra of the RS-effect are proportional to $\Omega_m\sigma_8^4$ (up to third order in perturbation theory), in contrast to the iSW-spectrum, which scales as $\Omega_m\sigma_8^2$. The dependence on the dark energy eos-parameter $w$ is weaker in the nonlinear effect and the shape of the spectrum (determined by $h$ and $n_s$) becomes less important because of the integrations over $\dd^3p$ carried out in perturbation theory. These arguments motivate the quantification of the RS-contamination of the iSW-spectrum, and their interference with the estimation of cosmological parameters.

\section{Angular power spectra}\label{sect_isw}
In summary, the line of sight integrals for the iSW-temperature perturbation $\tau^{(n)}$ and the galaxy density $\gamma^{(n)}$ in order $n$ read:
\begin{eqnarray}
\tau^{(n)} & = & \frac{3\Omega_m}{c}\int_0^{\chi_H}\dd\chi\: a^2H(a)\frac{\dd}{\dd a}\left(\frac{D^n_+}{a}\right)\:\varphi^{(n)},\\
\gamma^{(n)} & = & \int_0^{\chi_H}\dd\chi\: p(z)\frac{\dd z}{\dd\chi} D_+^n(\chi)\:\delta^{(n)},
\end{eqnarray}
where we have defined the dimensionless potential $\varphi^{(n)}\equiv\Delta^{-1}\delta^{(n)}/\chi_H^2$ from the inversion of the Poisson equation, rescaled with the square of the Hubble distance $\chi_H=c/H_0$ for convenience. Due to the linearity of the Newtonian Poisson-relation, the perturbative corrections in $\delta$ map directly onto the corrections in $\varphi$. The weighting functions
\begin{eqnarray}
W^{(n)}_\tau(\chi) & = & \frac{3\Omega_m}{c}a^2 H(a) \frac{\dd}{\dd a}\frac{D_+^n}{a},\\
W^{(n)}_\gamma(\chi) & = & p(z)\frac{\dd z}{\dd\chi} D_+^n(\chi),
\end{eqnarray}
can be identified, which allow the expressions for the angular cross spectra to be written in a compact notation, applying a Limber-projection \citep{1954ApJ...119..655L} in the flat-sky approximation.
\begin{equation}
C_{\tau\gamma}^{(11)}(\ell) = \int_0^{\chi_H}\frac{\dd\chi}{\chi^2}\: W_\tau^{(1)}(\chi)W_\gamma^{(1)}(\chi)P_{\delta\varphi}^{(11)}(\ell/\chi)
\end{equation}
\begin{equation}
C_{\tau\gamma}^{(13)}(\ell) = \int_0^{\chi_H}\frac{\dd\chi}{\chi^2}\: \left(W_\tau^{(1)}(\chi)W_\gamma^{(3)}(\chi)+W_\tau^{(3)}(\chi)W_\gamma^{(1)}(\chi)\right)P_{\delta\varphi}^{(13)}(\ell/\chi)
\end{equation}
\begin{equation}
C_{\tau\gamma}^{(22)}(\ell) = \int_0^{\chi_H}\frac{\dd\chi}{\chi^2}\: W_\tau^{(2)}(\chi)W_\gamma^{(2)}(\chi)P_{\delta\varphi}^{(22)}(\ell/\chi)
\end{equation}
with the cross-spectrum $P_{\delta\varphi}^{(ij)}(k) = P_{\delta\delta}^{(ij)}(k) / (\chi_H k)^2$. The expression for the spectrum $C_{\tau\gamma}^{(13)}(\ell)$ has been symmetrised. The angular auto-spectra of the temperature perturbation $\tau$ are given by:
\begin{equation}
C_{\tau\tau}^{(11)}(\ell) = \int_0^{\chi_H}\frac{\dd\chi}{\chi^2}\: W_\tau^{(1)}(\chi)^2P_{\varphi\varphi}^{(11)}(\ell/\chi)
\end{equation}
\begin{equation}
C_{\tau\tau}^{(13)}(\ell) = 2\int_0^{\chi_H}\frac{\dd\chi}{\chi^2}\: W_\tau^{(1)}(\chi)W_\tau^{(3)}(\chi)P_{\varphi\varphi}^{(13)}(\ell/\chi)
\end{equation}
\begin{equation}
C_{\tau\tau}^{(22)}(\ell) = \int_0^{\chi_H}\frac{\dd\chi}{\chi^2}\: W_\tau^{(2)}(\chi)^2P_{\varphi\varphi}^{(22)}(\ell/\chi)
\end{equation}
In analogy to $P_{\delta\varphi}^{(ij)}(k)$, the spectrum of the potential $\varphi$ is defined as $P_{\varphi\varphi}^{(ij)}(k) = P_{\delta\delta}^{(ij)}(k) / (\chi_H k)^4$. Finally, the spectra of the galaxy density $\gamma$ can be evaluated to be:
\begin{equation}
C_{\gamma\gamma}^{(11)}(\ell) = \int_0^{\chi_H}\frac{\dd\chi}{\chi^2}\: W_\gamma^{(1)}(\chi)^2P_{\delta\delta}^{(11)}(\ell/\chi)
\end{equation}
\begin{equation}
C_{\gamma\gamma}^{(13)}(\ell) = 2\int_0^{\chi_H}\frac{\dd\chi}{\chi^2}\: W_\gamma^{(1)}(\chi)W_\gamma^{(3)}(\chi) P_{\delta\delta}^{(13)}(\ell/\chi)
\end{equation}
\begin{equation}
C_{\gamma\gamma}^{(22)}(\ell) = \int_0^{\chi_H}\frac{\dd\chi}{\chi^2}\: W_\gamma^{(2)}(\chi)^2P_{\delta\delta}^{(22)}(\ell/\chi)
\end{equation}
Collecting all terms, the full spectra consists of one first and two second order contributions,
\begin{eqnarray}
C_{\tau\gamma}(\ell) & = & C_{\tau\gamma}^{(11)}(\ell) + C_{\tau\gamma}^{(22)}(\ell) + C_{\tau\gamma}^{(13)}(\ell),\\
C_{\tau\tau}(\ell) & = & C_{\tau\tau}^{(11)}(\ell) + C_{\tau\tau}^{(22)}(\ell) + C_{\tau\tau}^{(13)}(\ell),\vphantom{C_{\gamma\gamma}^{(22)}(\ell)}\\
C_{\gamma\gamma}(\ell) & = & C_{\gamma\gamma}^{(11)}(\ell) + C_{\gamma\gamma}^{(22)}(\ell) + C_{\gamma\gamma}^{(13)}(\ell).
\end{eqnarray}

\begin{figure}
\resizebox{\hsize}{!}{\includegraphics{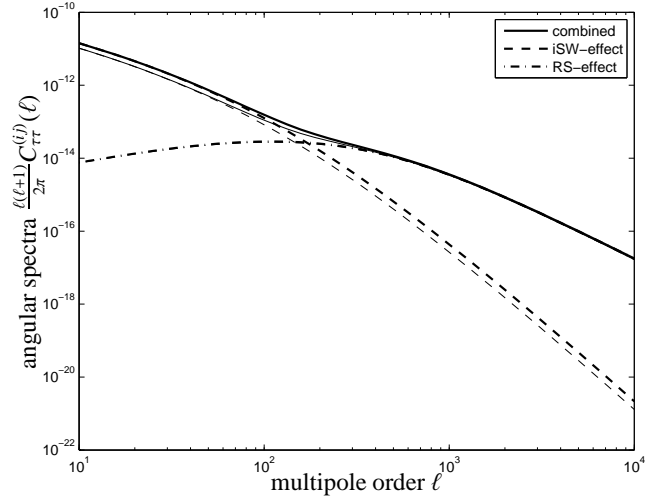}}
\caption{Angular iSW-spectrum $C_{\tau\tau}(\ell)$ of the iSW-effect (solid line), split up into the linear effect $C_{\tau\tau}^{(11)}(\ell)$ (dashed line) and the nonlinear RS-corrections $C_{\tau\tau}^{(22)}(\ell) + C_{\tau\tau}^{(13)}(\ell)$ (dash-dotted line). The plot compares spectra for $w$CDM with $w=-0.9$ (thick lines) with $\Lambda$CDM with $w=-1$ (thin lines).}
\label{fig_spectra_auto}
\end{figure}

\begin{figure}
\resizebox{\hsize}{!}{\includegraphics{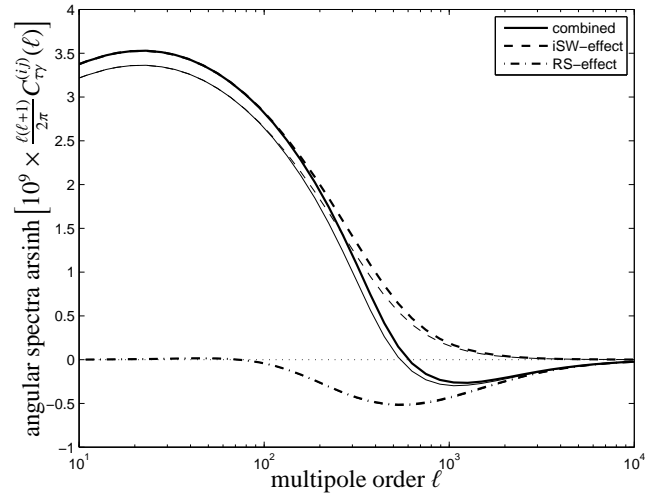}}
\caption{Angular iSW-cross spectrum $C_{\tau\gamma}(\ell)$ of the iSW-effect (solid line), split up into the linear effect $C_{\tau\gamma}^{(11)}(\ell)$ (dashed line) and the nonlinear RS-corrections $C_{\tau\gamma}^{(22)}(\ell) + C_{\tau\gamma}^{(13)}(\ell)$ (dash-dotted line), both for $w$CDM ($w=-0.9$, thick lines) and $\Lambda$CDM ($w=-1$, thin lines).}
\label{fig_spectra_cross}
\end{figure}

Figs.~\ref{fig_spectra_auto} and~\ref{fig_spectra_cross} give the iSW-auto and cross-spectra, respectively, split up into linear contributions and the two perturbative corrections. The iSW-auto spectrum is dominated on multipoles larger than $100$ and the cross-spectrum is suppressed by the negative correlation between iSW-effect and tracer density on similar scales, leading to a sign change of the cross-spectrum at $\ell\simeq500$, which confirms earlier perturbative and $n$-body results \citep{1996ApJ...460..549S, 2002PhRvD..65h3518C, 2008ApJ...676L..93N, 2008arXiv0809.4488C, 2009arXiv0905.2408S}, but using a different perturbation theory approach. $\mathrm{arsinh}(x)$ is equal to $x$ for $\left|x\right|\ll 1$ and $\propto\ln x$ for $\left| x\right|\gg 1$, which allows to show the logarithmic behaviour of $C_{\tau\gamma}(\ell)$ despite the sign change. Another interesting feature is the fact that the nonlinear effect is much less sensitive on the choice of cosmological parameters, in particular the dark energy equation of state parameter $w$, for which the RS-spectra depicted differ by about 5\%. The sign change and its sensitivity on $w$ is mostly driven by changes in $C_{\tau\gamma}^{(11)}(\ell)$. Fig.~\ref{figure_iswrs_logscale} gives the cross-spectrum in a logarithmic representation and shows that the anti-correlation between $\tau$ and $\gamma$ is a generic feature of nonlinearly evolving structures from angular scales of $\ell\simeq70$ on, but the linear effect shifts the anticorrelation scales to much higher multipole moments. The sign change can be easily explained by the fact that in linear structure formation potentials are constant or decay slowly, depending on cosmology, whereas in nonlinear structure formation the potentials grow fast, which manifests itself in the iSW-effect by causing temperature perturbations of opposite sign.

\section{statistical errors}\label{isw_fisher}
In this chapter, we recapitulate the estimation of statistical precision on parameters derived from angular iSW-spectra with Fisher matrices \citep{1997ApJ...480...22T}, and the accuracy of the parameter estimation with an extended Fisher formalism \citep{2007MNRAS.381.1347C,2007arXiv0710.5171A,2009MNRAS.392.1153T}.

\subsection{Fisher-matrix for the iSW-spectrum $C_{\tau\gamma}(\ell)$}
The Fisher matrix, which quantifies the decrease in likelihood if a model parameter $x_\mu$ moves away from the fiducial value, can be computed for a local Gaussian approximation to likelihood $\mathcal{L}\propto\exp(-\chi^2/2)$. The Fisher-matrix for the measurement of $C_{\tau\gamma}(\ell)$ is given by
\begin{equation}
F_{\mu\nu}^\mathrm{iSW} = \sum_{\ell=\ell_\mathrm{min}}^{\ell_\mathrm{max}}
\frac{\partial C_{\tau\gamma}(\ell)}{\partial x_\mu}
\mathrm{Cov}^{-1}\left(C_{\tau\gamma}(\ell),C_{\tau\gamma}(\ell)\right)
\frac{\partial C_{\tau\gamma}(\ell)}{\partial x_\nu}.
\end{equation}
We construct the Fisher-matrix $F_{\mu\nu}$ for $\Lambda$CDM as the fiducial cosmological model, with fiducial values for the parameters being $\Omega_m=0.25$, $\sigma_8=0.8$, $h=0.72$, $n_s=1$ and $w=-1$. Implicitly, we assume priors on spatial flatness, $\Omega_m+\Omega_\Lambda=1$ and neglect the weak dependence of the shape parameter on the baryon density $\Omega_b$. CMB-priors on the cosmological parameters are incorporated by adding the CMB Fisher matrix $F_{\mu\nu}^\mathrm{CMB}$,
\begin{equation}
F_{\mu\nu} = F_{\mu\nu}^\mathrm{iSW} + F_{\mu\nu}^\mathrm{CMB}.
\end{equation}

\subsection{Noise modelling}
In an actual observation, the iSW-power spectrum is modified by the intrinsic CMB-fluctuations, the instrumental noise and the beam as noise sources, assuming mutual uncorrelatedness of the individual contributions. The galaxy correlation function assumes a Poissonian noise term,
\begin{eqnarray}
\tilde{C}_{\tau\tau}(\ell) & = & C_{\tau\tau}(\ell) + C_\mathrm{CMB}(\ell) + w_T^{-1}B^{-2}(\ell),\\
\tilde{C}_{\gamma\gamma}(\ell) & = & C_{\gamma\gamma}(\ell) + \frac{1}{n},\label{eqn_obs_gg}
\end{eqnarray}
For PLANCK's noise levels the value $w_T^{-1}=(0.02\umu\mathrm{K})^2$ has been used, and the beam was assumed to be Gaussian, $B^{-2}(\ell)=\exp(\Delta\theta^2\:\ell(\ell+1))$, with a FWHM-width of $\Delta\theta = 7\farcm1$, corresponding to channels of PLANCK closest to the CMB-maximum at $\simeq160~\mathrm{GHz}$.

EUCLID is designed to survey the entire extragalactic sky and to cover the solid angle $\Delta\Omega=2\pi$, corresponding to $f_\mathrm{sky}=0.5$, yielding a total of $n=4.7\times10^8$ galaxies per steradian at a density of 40 galaxies per squared arcminute. The observed cross power spectra are unbiased estimates of the actual spectra,
\begin{equation}
\tilde{C}_{\tau\gamma}(\ell) = C_{\tau\gamma}(\ell),
\end{equation}
in the case of uncorrelated noise terms. We determine the spectrum $C_\mathrm{CMB}(\ell)$ of the primary CMB anisotropies with the CAMB code \citep{2000ApJ...538..473L}. The covariance of the spectrum $C_{\tau\gamma}(\ell)$ is given in terms of the observed spectra $\tilde{C}_{\tau\tau}(\ell)$, $\tilde{C}_{\tau\gamma}(\ell)$ and $\tilde{C}_{\gamma\gamma}(\ell)$ which follow directly from applying the Wick-theorem,
\begin{equation}
\mathrm{Cov}(C_{\tau\gamma},C_{\tau\gamma}) = 
\frac{1}{2\ell+1}\frac{1}{f_\mathrm{sky}}
\left[\tilde{C}_{\tau\gamma}^2(\ell) + \tilde{C}_{\tau\tau}(\ell)\tilde{C}_{\gamma\gamma}(\ell)\right].
\end{equation}
In all applications considered in this paper, PLANCK causes the dominating noise contribution in comparison to the Poisson noise in the galaxy number density given by EUCLID.

\subsection{Detectability of the RS-effect}
\begin{figure}
\resizebox{\hsize}{!}{\includegraphics{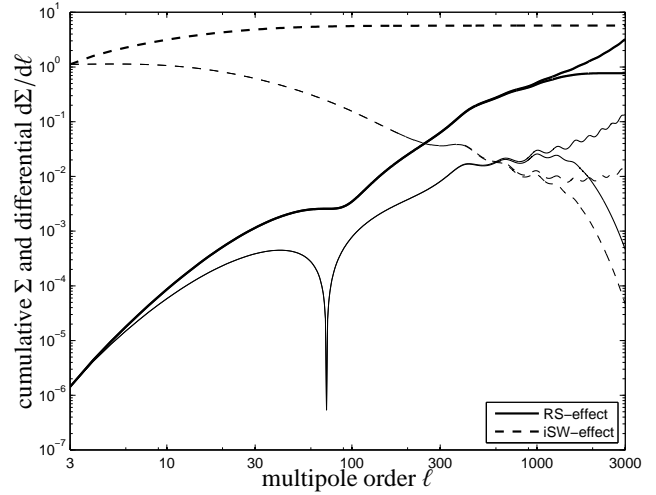}}
\caption{Signal to noise ratio of a measurement of the cross-spectrum $C_{\tau\gamma}(\ell)$, cumulative $\Sigma$ (thick lines) and differential $\dd\Sigma/\dd\ell$ (thin lines), for the linear iSW-effect (dashed line) and the nonlinear RS-effect (solid line). The plot compares the signal to noise ratio attainable with hypothetical cosmic-variance limited experiments (upper set of lines) with that reachable by combining PLANCK with EUCLID (lower set of lines).}
\label{fig_s2n}
\end{figure}

The signal to noise ratio $\Sigma$ of the cross-spectrum $C_{\tau\gamma}(\ell)$ reads:
\begin{equation}
\Sigma^2 = \sum_{\ell=\ell_\mathrm{min}}^{\ell_\mathrm{max}} \frac{C_{\tau\gamma}^2(\ell)}{\mathrm{Cov}(C_{\tau\gamma}(\ell))},
\end{equation}
for mutually uncorrelated modes as in the case of a full-sky observation. Fig.~\ref{fig_s2n} shows the signal to noise ratio $\Sigma$ of a measurement of $C_{\tau\gamma}(\ell)$ including the nonlinear contribution at high $\ell$. The figure suggests thay ideal cosmic variance limited experiments can in fact detect the RS-effect with a significance of $3.22\sigma$ (corresponding to a confidence of 0.998) integrating over all multipoles up to $\ell=3\times10^3$, and that this significance is reduced by the finite resolution and the noise of PLANCK to a mere $0.77\sigma$ (0.558 confidence). Thus, the signal to noise ratio of the nonlinear effect is roughly smaller by an order of magnitude compared to that of the linear iSW-effect. Apart from the increasing correlation noise at high $\ell$ it is the smallness of the spectrum around the cross-over scale which does not provide enough signal for a detection. Between $\ell=30$ and $\ell=100$ the cumulative signal to noise ratio stagnates, which is not included in the computation by \citet{2002PhRvD..65h3518C} as his perturbative approach is not able to reproduce the small values of $\dd\Sigma/\dd l$ due to the sign-change of $C_{\tau\gamma}(\ell)$ at $\ell\simeq70$. Despite the sensitivity of the cross-over scale on e.g. the dark energy equation of state parameter $w$ it would be very difficult to measure this scale as the signal to noise ratio of each multipole is about $10^{-3}$ and as the same knowledge on $w$  can be already derived from much smaller multipoles with sufficient accuracy.

\subsection{Parameter bounds and degeneracies}
\begin{figure*}
\vspace{0.5cm}
\resizebox{0.85\hsize}{!}{\includegraphics{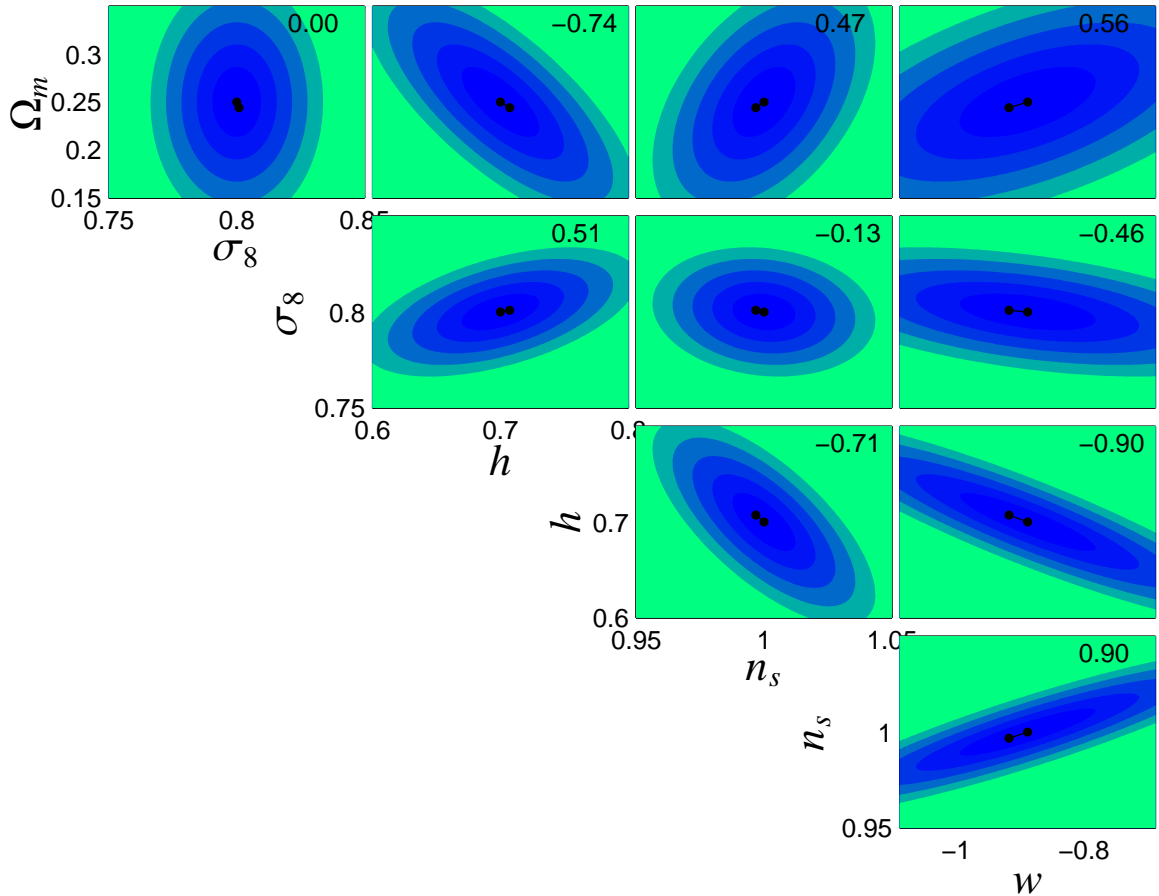}}
\vspace{0.5cm}
\caption{Constraints on the parameters $\Omega_m$, $\sigma_8$, $n_s$, $h$ and $w$, from the cross correlation of PLANCK with EUCLID. The ellipses correspond to $1\ldots5\sigma$ confidence regions. Additionally, the vectors $(\delta_\mu,\delta_\nu)$ indicate the bias in the estimation of the cosmological parameters due to the nonlinear contributions and have been {\em enlarged by a factor of 10}. The estimation bias was derived for a multipole range extending up to $\ell_\mathrm{max}=3\times10^3$. The number in the upper right corner of each panel gives the correlation coefficient $r_{\mu\nu}$. The constraints include a prior from the statistics $C_\mathrm{CMB}(\ell)$ of primary CMB temperature anisotropies on both the statistical and systematical error.}
\label{fig_fisher}
\end{figure*}

The $\chi^2$-function for a pair of parameters $(x_\mu,x_\nu)$ can be computed from the inverse $(F^{-1})_{\mu\nu}$ of the Fisher matrix,
\begin{equation}
\chi^2 = 
\left(
\begin{array}{c}
\Delta x_\mu \\
\Delta x_\nu
\end{array}
\right)^t
\left(
\begin{array}{cc}
(F^{-1})_{\mu\mu} & (F^{-1})_{\mu\nu} \\
(F^{-1})_{\nu\mu} & (F^{-1})_{\nu\nu}
\end{array}
\right)^{-1}
\left(
\begin{array}{c}
\Delta x_\mu \\
\Delta x_\nu
\end{array}
\label{eqn_chi2}
\right),
\end{equation}
where $\Delta x_\mu = x_\mu - x_\mu^{\Lambda\mathrm{CDM}}$. The correlation coefficient $r_{\mu\nu}$ is defined as
\begin{equation}
r_{\mu\nu} =
\frac{(F^{-1})_{\mu\nu}}{\sqrt{(F^{-1})_{\mu\mu}(F^{-1})_{\nu\nu}}},
\end{equation}
and describes the degree of dependence between the parameters $x_\mu$ and $x_\nu$ by assuming numerical values close to 0 for independent, and close to unity for strongly dependent parameters. The degeneracies between the cosmological parameters $\Omega_m$, $\sigma_8$, $h$, $n_s$ and $w$ estimated from the linear iSW-effect is shown in Fig.~\ref{fig_fisher}, along with the correlation coefficient from a measurement combining PLANCK and EUCLID data up to very high multipoles of $\ell=3000$, including a prior from CMB data.

\section{systematical errors}\label{isw_bias}
In this section, we quantify how the interpretation of the data with the pure iSW-spectrum affects the estimation of cosmological parameters, if in reality there are nonlinear RS-contributions at higher multipoles. Using this formalism, we seek to minimise the combined statistical error by finding an optimal angular scale $\ell_\mathrm{opt}$ down to which the iSW-measurement should be carried out. The nonlinear iSW-effect is the dominant contamination of the iSW-spectrum at intermediate multipoles, with the kinetic Sunyaev-Zel'dovich effect becoming important at multipoles above $\ell\simeq 10^3$. The parameter estimation bias formalism has been validated with Monte-Carlo Markov-chains and was found to be an excellent approximation for weak systematics \citep{2010MNRAS.tmp..433T}.

\subsection{Estimation bias formalism}
The angular iSW-spectrum $C_{\tau\gamma}(\ell)=C_\mathrm{iSW}(\ell)+C_\mathrm{RS}(\ell)$ can be separated into the linear part $C_\mathrm{iSW}(\ell)$ and an additive systematic $C_\mathrm{RS}(\ell)$ due to the nonlinear corrections,
\begin{equation}
C_\mathrm{iSW}(\ell) = C_{\tau\gamma}^{(11)}(\ell)
\end{equation}
\begin{equation}
C_\mathrm{RS}(\ell) = C_{\tau\gamma}^{(22)}(\ell) + C_{\tau\gamma}^{(13)}(\ell)
\end{equation}
Using these relations, we define the power spectrum of the true model $C_t(\ell)$ including nonlinear corrections,
\begin{equation}
C_t(\ell) = 
C_\mathrm{iSW}(\ell) + C_\mathrm{RS}(\ell) = 
C_{\tau\gamma}^{(11)}(\ell) + C_{\tau\gamma}^{(22)}(\ell) + C_{\tau\gamma}^{(13)}(\ell)
\end{equation}
as well as the spectrum of the false model $C_f(\ell)$, which neglects these RS-contributions,
\begin{equation}
C_f(\ell) = C_\mathrm{iSW}(\ell) = C_{\tau\gamma}^{(11)}(\ell),
\end{equation}
where the observed spectra $\tilde{C}_i(\ell)$ are unbiased estimators of the theoretical spectra $C_i(\ell)$ in each case, because of uncorrelated errors in each observational channel in the cross-correlation measurement method. The estimation of cosmological parameters is carried out from maximisation of the $\chi^2$-functionals of the two competing models,
\begin{eqnarray}
\chi_t^2 & = & \sum_{\ell=\ell_\mathrm{min}}^{\ell_\mathrm{max}} 
\frac{\left(\tilde{C}_t(\ell)-C_t(\ell)\right)^2}{\mathrm{Cov}\left[C_t(\ell),C_t(\ell)\right]},\quad\mathrm{and}\\
\chi_f^2 & = & \sum_{\ell=\ell_\mathrm{min}}^{\ell_\mathrm{max}} 
\frac{\left(\tilde{C}_t(\ell)-C_f(\ell)\right)^2}{\mathrm{Cov}\left[C_f(\ell),C_f(\ell)\right]},
\end{eqnarray}
i.e. the data is in reality described by $C_t(\ell)$ and, in the second case, fitted wrongly with $C_f(\ell)$ instead of $C_t(\ell)$. The best-fit parameters $\bmath{x}$ for each model can be derived by solving the equations $\bra\partial\chi^2/\partial x_\mu\ket = 0$ following from the respective $\chi^2$-functional. 

For deriving the distance $\bmath{x}_f-\bmath{x}_t$ between the best-fit values of the true and the false model, we expand the $\chi^2_f$ function at the best-fit position $\bmath{x}_t$ in a Taylor series \citep[see][]{2009MNRAS.392.1153T}
\begin{equation}
\chi_f^2(\bmath{x}_f) = 
\chi_f^2(\bmath{x}_t) + 
\sum_\mu\frac{\partial}{\partial x_\mu}\chi_f^2(\bmath{x}_t)\: \delta_\mu + 
\frac{1}{2}\sum_{\mu,\nu}\frac{\partial^2}{\partial x_\mu\partial x_\nu}\chi_f^2(\bmath{x}_t)\: \delta_\mu\delta_\nu,
\end{equation}
where the parameter estimation bias vector $\bdelta \equiv \bmath{x}_f-\bmath{x}_t$ was defined. The best-fit position $\bmath{x}_f$ of $\chi_f^2$ can be recovered by extremisation of the ensemble-averaged $\bra\chi_f^2\ket$, yielding
\begin{equation}
\left\bra\frac{\partial}{\partial x_\mu}\chi_f^2\right\ket_{\bmath{x}_t} = 
-\sum_\nu\left\bra\frac{\partial^2}{\partial x_\mu\partial x_\nu}\chi_f^2\right\ket_{\bmath{x}_t}\delta_\nu,
\end{equation}
which is a linear system of equations of the form 
\begin{equation}
\sum_\nu G_{\mu\nu}\delta_\nu = a_\mu\rightarrow \delta_\mu = \sum_\nu (G^{-1})_{\mu\nu}a_\nu,
\end{equation}
where the two quantities $G_{\mu\nu}$ and $a_\mu$ follow from the derivatives of the $\chi_f^2$-function, evaluated at $\bmath{x}_t$,
\begin{eqnarray}
G_{\mu\nu}^\mathrm{iSW} & \equiv & 
\sum_{\ell=\ell_\mathrm{min}}^{\ell_\mathrm{max}}\mathrm{Cov}^{-1}\left[\frac{\partial C_\mathrm{iSW}(\ell)}{\partial x_\mu}\frac{\partial C_\mathrm{iSW}(\ell)}{\partial x_\nu} - C_\mathrm{RS}(\ell)\frac{\partial^2 C_\mathrm{iSW}(\ell)}{\partial x_\mu\partial x_\nu}\right],\nonumber\\
a_{\mu} & \equiv &
\sum_{\ell=\ell_\mathrm{min}}^{\ell_\mathrm{max}} \mathrm{Cov}^{-1}\left[C_\mathrm{RS}(\ell)\frac{\partial C_\mathrm{iSW}(\ell)}{\partial x_\mu}\right].
\end{eqnarray}
The CMB priors can be incorporated by adding the Fisher-matrix $F_{\mu\nu}^\mathrm{CMB}$ to $G_{\mu\nu}$,
\begin{equation}
G_{\mu\nu} = G_{\mu\nu}^\mathrm{iSW} + F_{\mu\nu}^\mathrm{CMB},
\label{eqn_likelihood_combined}
\end{equation}
for independent iSW- and CMB-likelihoods. The biases in parameter estimation from the iSW-effect are depicted alongside the degeneracies in Fig.~\ref{fig_fisher}, for a maximum multipole order of $\ell_\mathrm{max}=3000$. The parameter estimation biases in the combined set of cosmological parameters are very small, due to the weakness of the RS-signal in comparison to that of the iSW-effect, and due to the strong prior from primary CMB fluctuations. Typical values for misestimates in cosmological parameters are of the order of $<0.1\sigma$, and are negligible in comparison to statistical errors.

\subsection{Contamination of the iSW-spectrum}
In this section we consider the application of the iSW-effect for providing independent constraints on individual cosmological parameters. If the iSW-likelihood is combined with the CMB-likelihood according to eqn.~(\ref{eqn_likelihood_combined}), the latter is by far dominating due to larger signal to noise ratio. Although the signal strength of the iSW-effect is not enough for fully constraining a standard dark energy cosmology with five or more parameters, it is sufficient to place competitive bounds on single cosmological parameters. Therefore, we define the conditional systematical error $\sigma_\mu = 1/\sqrt{F_{\mu\mu}}$ and the systematical error $b_\mu$ on a single parameter $x_\mu$ while all other parameters are assumed to coincide exactly with their fiducial values.

\begin{figure}
\resizebox{\hsize}{!}{\includegraphics{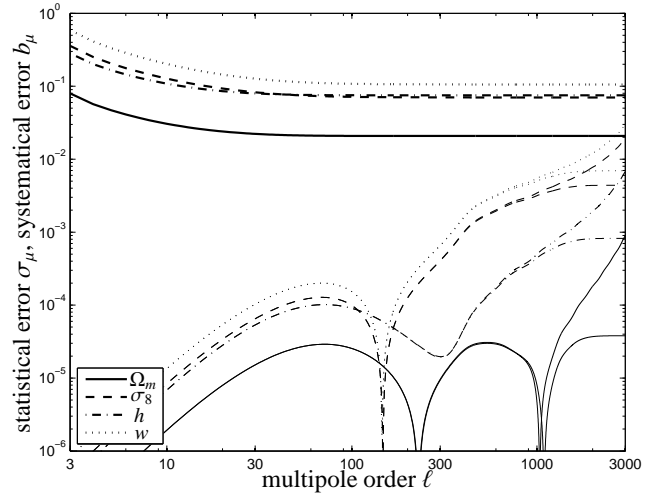}}
\caption{Conditional statistical errors $\sigma_\mu(\ell_\mathrm{max})$ (thick lines) and systematical errors $b_\mu(\ell_\mathrm{max})$ (thin lines) for individual cosmological parameters: $\Omega_m$ (solid line), $\sigma_8$ (dashed line), $h$ (dash-dotted line) and $w$ (dotted line) as a function of maximum multipole order $\ell_\mathrm{max}$. No prior information is used in the computation of the errors. The plot compares a cosmic variance limited measurement (upper lines) with the cross correlation of PLANCK with EUCLID (lower lines).}
\label{figure_opt_multipole}
\end{figure}

At low multipoles $\ell_\mathrm{max}$ the error budget will be dominated by statistics, while the systematics due to the nonlinear contributions are negligible. Conversely, the extention of the computation to higher multipoles $\ell_\mathrm{max}$ will reduce the statistical error, but the RS-contributions will start to deteriorate the parameter accuracy. Fig.~\ref{figure_opt_multipole} depicts the individual conditional statistical and systematical errors as a function of maximum multipole order $\ell_\mathrm{max}$. In comparison to the statistical errors on parameters derived with the iSW-spectrum, which are monotonically decreasing, the parameter estimation biases due to RS-contributions have a more complicated behaviour with multipole order $\ell$, but remain always small in comparison to the statistical error by more than one order of magnitude, for both cosmic variance dominated experiments and the combination of PLANCK with EUCLID. The worst case is the constraint on $w$ in a cosmic variance limited experiment, where the systematic error amounts to 20\% of that of the statistical error. There are certain scales at which the systematical errors are very small, namely as they change their signs, in agreement with changing parameter degeneracies on different angular scales.

\section{Summary}\label{sect_summary}
The topic of this paper is an investigation of the contamination due to nonlinearly evolving structures on the linear iSW-effect, and the consequent parameter estimation biases.
\begin{enumerate}
\item{The angular spectrum of the Rees-Sciama effect was computed in third order perturbation theory. The spectrum $C_{\tau\tau}(\ell)$ of the RS-effect starts dominating that of the linear integrated Sachs-Wolfe effect from multipoles of $\ell\simeq100$ on. In particular the cross-spectrum $C_{\tau\gamma}(\ell)$ shows a sign change suggesting that the CMB temperature is anticorrelated with the galaxy density on nonlinear scales. This scale bears some sensitivity on $w$ and is shifted from $\ell\simeq70$, where the sign change occurs in the spectrum of the nonlinear effect to $\ell\simeq500$ for the combination of the linear and nonlinear effect. The sensitivity of the RS-effect on the dark energy model itself is rather weak.}
\item{By combining the PLANCK and EUCLID data sets one can measure the nonlinear RS-effect with a significance of $0.77\sigma$ out to multipoles of $\ell=3000$, where the most important limitations are cosmic variance and PLANCK's instrumental noise. An ideal experiment only subjected to cosmic variance would be able to detect the effect with $3.22\sigma$, but at higher multipoles, confusion with the kinetic Sunyaev-Zel'dovich effect would occur. Measurements of the angular scale on which the sign change occurs are almost impossible as the signal to noise ratio on these scales is $\simeq10^{-3}$. In summary, the significance of the RS-effect is smaller to that of the iSW-effect by a factor of $\sim10$ for the combination of PLANCK with EUCLID, which reaches $\sim7\sigma$.}
\item{If constraints on cosmological parameters are derived from the linear iSW-effect and if contributions from the RS-effect are neglected, the induced parameter estimation biases are smaller than the statistical errors by one order of magnitude because of the smallness of the RS-effect in comparison to cosmic variance induced into the measurement by primary CMB fluctuations and because of the strong prior used. If the iSW-effect is used to constrain individual cosmological parameters without using a CMB-prior, a similar result still applies. Therefore, the RS-effect is negligible as a systematic in comparison to other systematics that have been discussed in the literature and which have a more pronounced effect on cosmogical parameters, e.g. redshift errors due to peculiar motion of the tracer galaxies \citep{2009arXiv0902.1759R}, weak lensing on the tracer population and galaxy magnification bias \citep{2007PhRvD..75d3519L}, bias evolution of the tracer population \citep{2008MNRAS.386.2161R, 2009arXiv0903.4288S}, and contributions due to the kinetic Sunyaev-Zel'dovich effect from reionisation \citep{2007MNRAS.381..819G}.}
\end{enumerate}
Given the estimates that both the spectrum and the bispectrum of the RS-effect are only detectable with significances of $\sim0.8\sigma$ casts doubt on the detectability of this effect in a statistical way, and emphasises the importance of alternative approaches such as stacking methods \citep{2008ApJ...683L..99G}.

\section*{Acknowledgements}
We would like to thank Nabila Aghanim, Marian Douspis, Carlos Hern{\'a}ndez-Monteagudo and Patricio Vielva for valuable comments. BMS would like to thank Subha Majumdar for his hospitality at IUCAA/Pune, where we started working on this paper. Marian Douspis provided a CMB Fisher matrix, Nicolas Taburet made his parameter estimation bias formalism available and Gero J{\"u}rgens shared his expertise on perturbation theory, for which we are very grateful. This work was supported by the German Research Foundation (DFG) within the framework of the excellence initiative through the Graduate School of Fundamental Physics in Heidelberg.

\bibliography{bibtex/aamnem,bibtex/references}

\begin{thebibliography}{}

\bibitem[\protect\citeauthoryear{{Abramowitz} \& {Stegun}}{{Abramowitz} \&
  {Stegun}}{1972}]{1972hmf..book.....A}
{Abramowitz} M.,  {Stegun} I.~A.,  1972, {Handbook of Mathematical Functions}.
Handbook of Mathematical Functions, New York: Dover, 1972

\bibitem[\protect\citeauthoryear{{Amara} \& {Refregier}}{{Amara} \&
  {Refregier}}{2007}]{2007arXiv0710.5171A}
{Amara} A.,  {Refregier} A.,  2007, ArXiv 0710.5171, 710

\bibitem[\protect\citeauthoryear{{Bardeen}, {Bond}, {Kaiser} \&
  {Szalay}}{{Bardeen} et~al.}{1986}]{1986ApJ...304...15B}
{Bardeen} J.~M.,  {Bond} J.~R.,  {Kaiser} N.,    {Szalay} A.~S.,  1986, \apj,
  304, 15

\bibitem[\protect\citeauthoryear{{Bernardeau}, {Colombi}, {Gazta{\~n}aga} \&
  {Scoccimarro}}{{Bernardeau} et~al.}{2002}]{2002PhR...367....1B}
{Bernardeau} F.,  {Colombi} S.,  {Gazta{\~n}aga} E.,    {Scoccimarro} R.,
  2002, \physrep, 367, 1

\bibitem[\protect\citeauthoryear{{Boughn} \& {Crittenden}}{{Boughn} \&
  {Crittenden}}{2004}]{2004Natur.427...45B}
{Boughn} S.,  {Crittenden} R.,  2004, \nat, 427, 45

\bibitem[\protect\citeauthoryear{{Cabr{\'e}}, {Fosalba}, {Gazta{\~n}aga} \&
  {Manera}}{{Cabr{\'e}} et~al.}{2007}]{2007MNRAS.381.1347C}
{Cabr{\'e}} A.,  {Fosalba} P.,  {Gazta{\~n}aga} E.,    {Manera} M.,  2007,
  \mnras, 381, 1347

\bibitem[\protect\citeauthoryear{{Cabr{\'e}}, {Gazta{\~n}aga}, {Manera},
  {Fosalba} \& {Castander}}{{Cabr{\'e}} et~al.}{2006}]{2006MNRAS.372L..23C}
{Cabr{\'e}} A.,  {Gazta{\~n}aga} E.,  {Manera} M.,  {Fosalba} P.,
  {Castander} F.,  2006, \mnras, 372, L23

\bibitem[\protect\citeauthoryear{{Cai}, {Cole}, {Jenkins} \& {Frenk}}{{Cai}
  et~al.}{2010}]{2010arXiv1003.0974C}
{Cai} Y.,  {Cole} S.,  {Jenkins} A.,    {Frenk} C.~S.,  2010, ArXiv 1003.0974

\bibitem[\protect\citeauthoryear{{Cai}, {Cole}, {Jenkins} \& {Frenk}}{{Cai}
  et~al.}{2008}]{2008arXiv0809.4488C}
{Cai} Y.-C.,  {Cole} S.,  {Jenkins} A.,    {Frenk} C.,  2008, ArXiv 0809.4488

\bibitem[\protect\citeauthoryear{{Cooray}}{{Cooray}}{2002}]{2002PhRvD..65h3518%
C}
{Cooray} A.,  2002, \prd, 65, 083518

\bibitem[\protect\citeauthoryear{{Crittenden} \& {Turok}}{{Crittenden} \&
  {Turok}}{1996}]{1996PhRvL..76..575C}
{Crittenden} R.~G.,  {Turok} N.,  1996, Physical Review Letters, 76, 575

\bibitem[\protect\citeauthoryear{{Douspis}, {Castro}, {Caprini} \&
  {Aghanim}}{{Douspis} et~al.}{2008}]{2008arXiv0802.0983D}
{Douspis} M.,  {Castro} P.~G.,  {Caprini} C.,    {Aghanim} N.,  2008, ArXiv
  0802.0983, 802

\bibitem[\protect\citeauthoryear{{Fosalba}, {Gazta{\~n}aga} \&
  {Castander}}{{Fosalba} et~al.}{2003}]{2003ApJ...597L..89F}
{Fosalba} P.,  {Gazta{\~n}aga} E.,    {Castander} F.~J.,  2003, \apjl, 597, L89

\bibitem[\protect\citeauthoryear{{Gazta{\~n}aga}, {Manera} \&
  {Multam{\"a}ki}}{{Gazta{\~n}aga} et~al.}{2006}]{2006MNRAS.365..171G}
{Gazta{\~n}aga} E.,  {Manera} M.,    {Multam{\"a}ki} T.,  2006, \mnras, 365,
  171

\bibitem[\protect\citeauthoryear{{Giannantonio} \& {Crittenden}}{{Giannantonio}
  \& {Crittenden}}{2007}]{2007MNRAS.381..819G}
{Giannantonio} T.,  {Crittenden} R.,  2007, \mnras, 381, 819

\bibitem[\protect\citeauthoryear{{Giannantonio}, {Crittenden}, {Nichol},
  {Scranton}, {Richards}, {Myers}, {Brunner}, {Gray}, {Connolly} \&
  {Schneider}}{{Giannantonio} et~al.}{2006}]{2006PhRvD..74f3520G}
{Giannantonio} T.,  {Crittenden} R.~G.,  {Nichol} R.~C.,  {Scranton} R.,
  {Richards} G.~T.,  {Myers} A.~D.,  {Brunner} R.~J.,  {Gray} A.~G.,
  {Connolly} A.~J.,    {Schneider} D.~P.,  2006, \prd, 74, 063520

\bibitem[\protect\citeauthoryear{{Giannantonio}, {Scranton}, {Crittenden},
  {Nichol}, {Boughn}, {Myers} \& {Richards}}{{Giannantonio}
  et~al.}{2008}]{2008arXiv0801.4380G}
{Giannantonio} T.,  {Scranton} R.,  {Crittenden} R.~G.,  {Nichol} R.~C.,
  {Boughn} S.~P.,  {Myers} A.~D.,    {Richards} G.~T.,  2008, ArXiv 0801.4380,
  801

\bibitem[\protect\citeauthoryear{{Goldberg} \& {Spergel}}{{Goldberg} \&
  {Spergel}}{1999}]{1999PhRvD..59j3002G}
{Goldberg} D.~M.,  {Spergel} D.~N.,  1999, \prd, 59, 103002

\bibitem[\protect\citeauthoryear{{Granett}, {Neyrinck} \& {Szapudi}}{{Granett}
  et~al.}{2008}]{2008ApJ...683L..99G}
{Granett} B.~R.,  {Neyrinck} M.~C.,    {Szapudi} I.,  2008, \apjl, 683, L99

\bibitem[\protect\citeauthoryear{{Hu} \& {Sugiyama}}{{Hu} \&
  {Sugiyama}}{1994}]{1994PhRvD..50..627H}
{Hu} W.,  {Sugiyama} N.,  1994, \prd, 50, 627

\bibitem[\protect\citeauthoryear{{Lewis}, {Challinor} \& {Lasenby}}{{Lewis}
  et~al.}{2000}]{2000ApJ...538..473L}
{Lewis} A.,  {Challinor} A.,    {Lasenby} A.,  2000, \apj, 538, 473

\bibitem[\protect\citeauthoryear{{Limber}}{{Limber}}{1954}]{1954ApJ...119..655%
L}
{Limber} D.~N.,  1954, \apj, 119, 655

\bibitem[\protect\citeauthoryear{{Linder} \& {Jenkins}}{{Linder} \&
  {Jenkins}}{2003}]{2003MNRAS.346..573L}
{Linder} E.~V.,  {Jenkins} A.,  2003, \mnras, 346, 573

\bibitem[\protect\citeauthoryear{{Loverde}, {Hui} \& {Gazta{\~n}aga}}{{Loverde}
  et~al.}{2007}]{2007PhRvD..75d3519L}
{Loverde} M.,  {Hui} L.,    {Gazta{\~n}aga} E.,  2007, \prd, 75, 043519

\bibitem[\protect\citeauthoryear{{Maturi}, {Dolag}, {Waelkens}, {Springel} \&
  {En{\ss}lin}}{{Maturi} et~al.}{2007}]{2007A&A...476...83M}
{Maturi} M.,  {Dolag} K.,  {Waelkens} A.,  {Springel} V.,    {En{\ss}lin} T.,
  2007, \aap, 476, 83

\bibitem[\protect\citeauthoryear{{McEwen}, {Vielva}, {Hobson},
  {Mart{\'{\i}}nez-Gonz{\'a}lez} \& {Lasenby}}{{McEwen}
  et~al.}{2007}]{2007MNRAS.376.1211M}
{McEwen} J.~D.,  {Vielva} P.,  {Hobson} M.~P.,  {Mart{\'{\i}}nez-Gonz{\'a}lez}
  E.,    {Lasenby} A.~N.,  2007, \mnras, 376, 1211

\bibitem[\protect\citeauthoryear{{Mollerach}, {Gangui}, {Lucchin} \&
  {Matarrese}}{{Mollerach} et~al.}{1995}]{1995ApJ...453....1M}
{Mollerach} S.,  {Gangui} A.,  {Lucchin} F.,    {Matarrese} S.,  1995, \apj,
  453, 1

\bibitem[\protect\citeauthoryear{{Munshi}, {Souradeep} \&
  {Starobinsky}}{{Munshi} et~al.}{1995}]{1995ApJ...454..552M}
{Munshi} D.,  {Souradeep} T.,    {Starobinsky} A.~A.,  1995, \apj, 454, 552

\bibitem[\protect\citeauthoryear{{Nishizawa}, {Komatsu}, {Yoshida}, {Takahashi}
  \& {Sugiyama}}{{Nishizawa} et~al.}{2008}]{2008ApJ...676L..93N}
{Nishizawa} A.~J.,  {Komatsu} E.,  {Yoshida} N.,  {Takahashi} R.,    {Sugiyama}
  N.,  2008, \apjl, 676, L93

\bibitem[\protect\citeauthoryear{{Nolta}, {Wright}, {Page}, {Bennett},
  {Halpern}, {Hinshaw}, {Jarosik}, {Kogut}, {Limon}, {Meyer}, {Spergel},
  {Tucker} \& {Wollack}}{{Nolta} et~al.}{2004}]{2004ApJ...608...10N}
{Nolta} M.~R.,  {Wright} E.~L.,  {Page} L.,  {Bennett} C.~L.,  {Halpern} M.,
  {Hinshaw} G.,  {Jarosik} N.,  {Kogut} A.,  {Limon} M.,  {Meyer} S.~S.,
  {Spergel} D.~N.,  {Tucker} G.~S.,    {Wollack} E.,  2004, \apj, 608, 10

\bibitem[\protect\citeauthoryear{{Padmanabhan}, {Hirata}, {Seljak}, {Schlegel},
  {Brinkmann} \& {Schneider}}{{Padmanabhan} et~al.}{2005}]{2005PhRvD..72d3525P}
{Padmanabhan} N.,  {Hirata} C.~M.,  {Seljak} U.,  {Schlegel} D.~J.,
  {Brinkmann} J.,    {Schneider} D.~P.,  2005, \prd, 72, 043525

\bibitem[\protect\citeauthoryear{{Pietrobon}, {Balbi} \&
  {Marinucci}}{{Pietrobon} et~al.}{2006}]{2006PhRvD..74d3524P}
{Pietrobon} D.,  {Balbi} A.,    {Marinucci} D.,  2006, \prd, 74, 043524

\bibitem[\protect\citeauthoryear{{Raccanelli}, {Bonaldi}, {Negrello},
  {Matarrese}, {Tormen} \& {de Zotti}}{{Raccanelli}
  et~al.}{2008}]{2008MNRAS.386.2161R}
{Raccanelli} A.,  {Bonaldi} A.,  {Negrello} M.,  {Matarrese} S.,  {Tormen} G.,
    {de Zotti} G.,  2008, \mnras, 386, 2161

\bibitem[\protect\citeauthoryear{{Rassat}}{{Rassat}}{2009}]{2009arXiv0902.1759%
R}
{Rassat} A.,  2009, ArXiv 0902.1759

\bibitem[\protect\citeauthoryear{{Rassat}, {Land}, {Lahav} \&
  {Abdalla}}{{Rassat} et~al.}{2007}]{2007MNRAS.377.1085R}
{Rassat} A.,  {Land} K.,  {Lahav} O.,    {Abdalla} F.~B.,  2007, \mnras, 377,
  1085

\bibitem[\protect\citeauthoryear{{Rees} \& {Sciama}}{{Rees} \&
  {Sciama}}{1968}]{rees_sciama_orig}
{Rees} M.~J.,  {Sciama} D.~W.,  1968, Nature, 217, 511

\bibitem[\protect\citeauthoryear{{Refregier} \& {the DUNE
  collaboration}}{{Refregier} \& {the DUNE
  collaboration}}{2008}]{2008arXiv0802.2522R}
{Refregier} A.,  {the DUNE collaboration} 2008, ArXiv 0802.2522, 802

\bibitem[\protect\citeauthoryear{{Sachs} \& {Wolfe}}{{Sachs} \&
  {Wolfe}}{1967}]{1967ApJ...147...73S}
{Sachs} R.~K.,  {Wolfe} A.~M.,  1967, \apj, 147, 73

\bibitem[\protect\citeauthoryear{{Sahni} \& {Coles}}{{Sahni} \&
  {Coles}}{1995}]{1995PhR...262....1S}
{Sahni} V.,  {Coles} P.,  1995, \physrep, 262, 1

\bibitem[\protect\citeauthoryear{{Sch{\"a}fer}}{{Sch{\"a}fer}}{2008}]{2008MNRA%
S.388.1394S}
{Sch{\"a}fer} B.~M.,  2008, \mnras, 388, 1394

\bibitem[\protect\citeauthoryear{{Sch{\"a}fer} \& {Bartelmann}}{{Sch{\"a}fer}
  \& {Bartelmann}}{2006}]{2006MNRAS.369..425S}
{Sch{\"a}fer} B.~M.,  {Bartelmann} M.,  2006, \mnras, 369, 425

\bibitem[\protect\citeauthoryear{{Sch{\"a}fer}, {Douspis} \&
  {Aghanim}}{{Sch{\"a}fer} et~al.}{2009}]{2009arXiv0903.4288S}
{Sch{\"a}fer} B.~M.,  {Douspis} M.,    {Aghanim} N.,  2009, ArXiv 0903.4288

\bibitem[\protect\citeauthoryear{{Seljak}}{{Seljak}}{1996}]{1996ApJ...460..549%
S}
{Seljak} U.,  1996, \apj, 460, 549

\bibitem[\protect\citeauthoryear{{Smail}, {Hogg}, {Blandford}, {Cohen}, {Edge}
  \& {Djorgovski}}{{Smail} et~al.}{1995}]{1995MNRAS.277....1S}
{Smail} I.,  {Hogg} D.~W.,  {Blandford} R.,  {Cohen} J.~G.,  {Edge} A.~C.,
  {Djorgovski} S.~G.,  1995, \mnras, 277, 1

\bibitem[\protect\citeauthoryear{{Smith}, {Hernandez-Monteagudo} \&
  {Seljak}}{{Smith} et~al.}{2009}]{2009arXiv0905.2408S}
{Smith} R.~E.,  {Hernandez-Monteagudo} C.,    {Seljak} U.,  2009, ArXiv
  0905.2408

\bibitem[\protect\citeauthoryear{{Smith}, {Peacock}, {Jenkins}, {White},
  {Frenk}, {Pearce}, {Thomas}, {Efstathiou} \& {Couchman}}{{Smith}
  et~al.}{2003}]{2003MNRAS.341.1311S}
{Smith} R.~E.,  {Peacock} J.~A.,  {Jenkins} A.,  {White} S.~D.~M.,  {Frenk}
  C.~S.,  {Pearce} F.~R.,  {Thomas} P.~A.,  {Efstathiou} G.,    {Couchman}
  H.~M.~P.,  2003, \mnras, 341, 1311

\bibitem[\protect\citeauthoryear{{Spergel} \& {Goldberg}}{{Spergel} \&
  {Goldberg}}{1999}]{1999PhRvD..59j3001S}
{Spergel} D.~N.,  {Goldberg} D.~M.,  1999, \prd, 59, 103001

\bibitem[\protect\citeauthoryear{{Taburet}, {Aghanim}, {Douspis} \&
  {Langer}}{{Taburet} et~al.}{2009}]{2009MNRAS.392.1153T}
{Taburet} N.,  {Aghanim} N.,  {Douspis} M.,    {Langer} M.,  2009, \mnras, 392,
  1153

\bibitem[\protect\citeauthoryear{{Taburet}, {Douspis} \& {Aghanim}}{{Taburet}
  et~al.}{2010}]{2010MNRAS.tmp..433T}
{Taburet} N.,  {Douspis} M.,    {Aghanim} N.,  2010, \mnras, pp 433--+

\bibitem[\protect\citeauthoryear{{Tegmark}, {Taylor} \& {Heavens}}{{Tegmark}
  et~al.}{1997}]{1997ApJ...480...22T}
{Tegmark} M.,  {Taylor} A.~N.,    {Heavens} A.~F.,  1997, \apj, 480, 22

\bibitem[\protect\citeauthoryear{{Tuluie}, {Laguna} \& {Anninos}}{{Tuluie}
  et~al.}{1996}]{1996ApJ...463...15T}
{Tuluie} R.,  {Laguna} P.,    {Anninos} P.,  1996, \apj, 463, 15

\bibitem[\protect\citeauthoryear{{Turner} \& {White}}{{Turner} \&
  {White}}{1997}]{1997PhRvD..56.4439T}
{Turner} M.~S.,  {White} M.,  1997, \prd, 56, 4439

\bibitem[\protect\citeauthoryear{{Vielva}, {Mart{\'{\i}}nez-Gonz{\'a}lez} \&
  {Tucci}}{{Vielva} et~al.}{2006}]{2006MNRAS.365..891V}
{Vielva} P.,  {Mart{\'{\i}}nez-Gonz{\'a}lez} E.,    {Tucci} M.,  2006, \mnras,
  365, 891

\bibitem[\protect\citeauthoryear{{Wang} \& {Steinhardt}}{{Wang} \&
  {Steinhardt}}{1998}]{1998ApJ...508..483W}
{Wang} L.,  {Steinhardt} P.~J.,  1998, \apj, 508, 483

\end{thebibliography}
\bibliographystyle{mn2e}

\appendix

\section{Nonlinear corrections in perturbation theory}
\begin{figure}
\resizebox{\hsize}{!}{\includegraphics{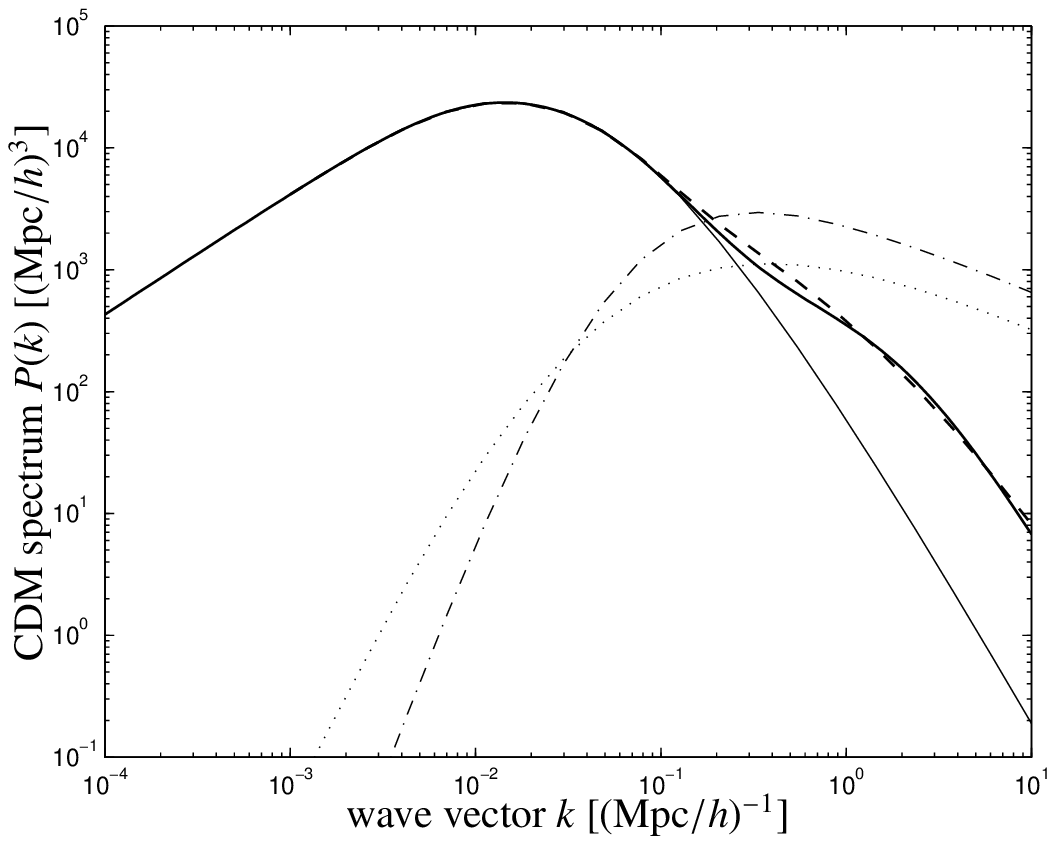}}
\caption{Nonlinear corrections to the linear CDM spectrum $P^{(11)}_{\delta\delta}(k)$ (thin solid line) of the density field in perturbation theory, at the current cosmic epoch: $P^{(22)}_{\delta\delta}(k)$ (thin dash-dotted line), the modulus of $P^{(13)}_{\delta\delta}(k)$ (thin dotted line), and their sum $P_{\delta\delta}(k)=P^{(11)}_{\delta\delta}(k)+P^{(22)}_{\delta\delta}(k)+2P^{(13)}_{\delta\delta}(k)$ (thick dashed line). The fit to the nonlinearly evolved spectrum $P_{\delta\delta}(k)$ from $n$-body data proposed by \citet{2003MNRAS.341.1311S} is given in comparison (thick solid line).}
\label{figure_nonlinear}
\end{figure}

Fig.~\ref{figure_nonlinear} illustrates the validity of the perturbative corrections to $P(k)$ due to nonlinear growth, by comparison to the result from $n$-body data \citep{2003MNRAS.341.1311S}. Third order perturbation theory is able to describe the increase in fluctuation amplitude due to nonlinear structure formation down to very small scales. One notices a deviation between the n-body result and the perturbation theory amounting to about 20\% in the transition region at a few inverse Mpc, corresponding to angular scales of $\ell\simeq300$, if most of the iSW-signal in the cross-correlation function arises at a comoving redshift of $\chi=2~\mathrm{Gpc}/h$, i.e. the maximum of the redshift distribution $p(z)\dd z$ used in this work. The higher orders beyond 3 in perturbation theory would correct the difference to the $n$-body result, and it should be kept in mind that the simulation on which the description by \citet{2003MNRAS.341.1311S} is based uses slightly different cosmological parameters, most notably higher $\Omega_m$ and $\sigma_8$.

\begin{figure}
\resizebox{\hsize}{!}{\includegraphics{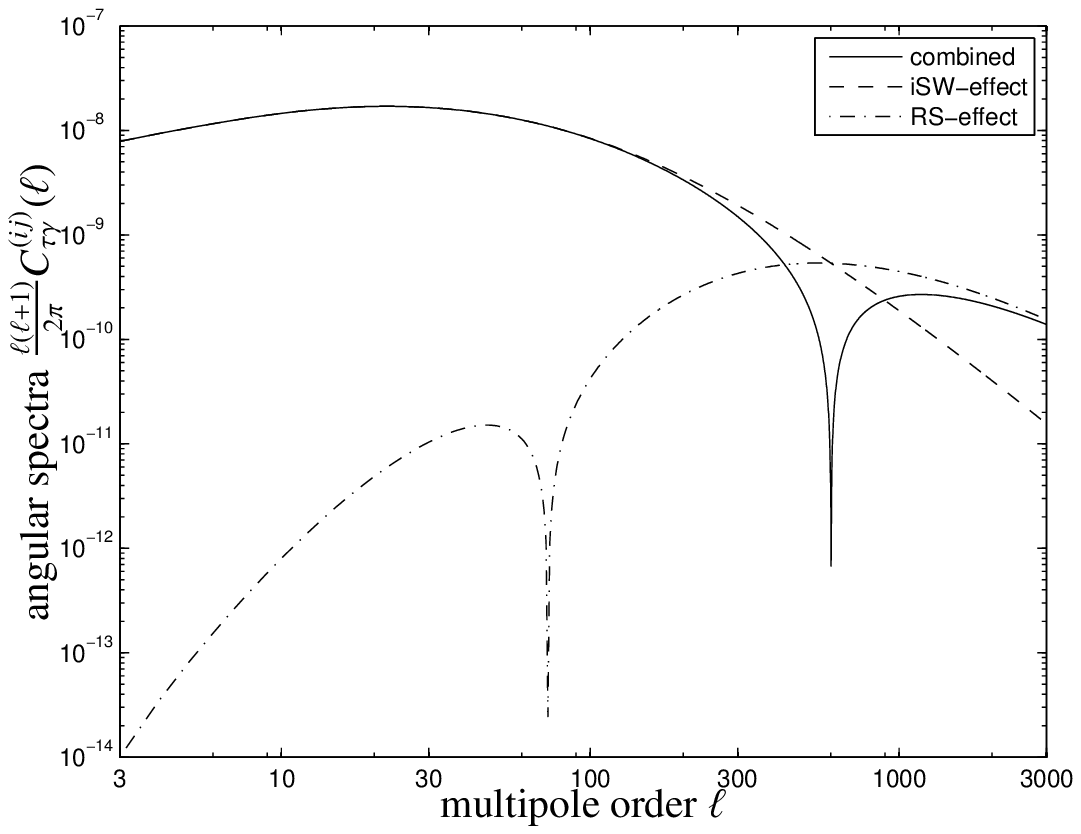}}
\caption{Angular iSW-cross spectrum $C_{\tau\gamma}(\ell)$ of the iSW-effect (solid line), split up into the linear effect $C_{\tau\gamma}^{(11)}(\ell)$ (dashed line) and the nonlinear RS-corrections $C_{\tau\gamma}^{(22)}(\ell) + C_{\tau\gamma}^{(13)}(\ell)$ (dash-dotted line) in logarithmic representation.}
\label{figure_iswrs_logscale}
\end{figure}

The remarkable behaviour of the nonlinear effect to cause an anticorrelation between the CMB and the tracer density is shown again in Fig.~\ref{figure_iswrs_logscale}, in a logarithmic representation. 

\bsp

\label{lastpage}

\end{document}